# Spatiotemporal topological phase transitions in photonic spacetime crystals

Zebin Zhu[1†], Bolun Huang[1†], Siqi Xu[2†], Jingming Chen[1], Yan Meng[3], Zhenxiao Zhu[4], Xiang Xi[3], Zhen Gao[1]*

[1]State Key Laboratory of Optical Fiber and Cable Manufacturing Technology, Department of Electronic and Electrical Engineering, Guangdong Key Laboratory of Integrated Optoelectronics Intellisense, Southern University of Science and Technology, Shenzhen, China

[2]Department of Electrical Engineering and Computer Science, Massachusetts Institute of Technology, Cambridge, MA, USA

[3]School of Electrical Engineering and Intelligentization, Dongguan University of Technology, Dongguan, China

[4]Frontier Interdisciplinary Domain, Beijing Institute of Technology, Zhuhai, China

[†]These authors contributed equally to this work.

[*]Corresponding author. Email: gaoz@sustech.edu.cn (Z.G.)

**Topological phase transitions have played a central role in topological physics. However, such transitions have so far been restricted to spatial or temporal crystals. Here, we transcend this conventional framework and report, for the first time, spatiotemporal topological phase transitions in photonic spacetime crystals — structures that are periodically modulated in both space and time. In a genuine photonic spacetime crystal composed of a dynamically modulated transmission-line metamaterial, we theoretically propose and experimentally demonstrate complete spatiotemporal topological phase transitions, characterized by the closing and reopening of both energy and momentum band gaps, along with changes in spatiotemporal topological invariants and topological phases. Furthermore, we directly observe a spatiotemporal, topologically localized state that exhibits causality-governed excitation and robustness to spatiotemporal disorders. Our findings reveal the interplay among space, time, and topology, establishing a unified framework that provides a comprehensive picture of the emerging topological spacetime physics and opening new avenues for robust spatiotemporal topological wave manipulations.**

## Introduction

The discovery of topological phases and topological phase transitions has profoundly reshaped modern condensed matter physics over the past decades[1]. To date, their influence has expanded into diverse physical disciplines, including photonics[2], acoustics[3], phononics[4], thermotics[5], and electrical circuits[6], opening new avenues for robust topological devices that are immune to disorder and structural defects. Conventionally, topological phases and topological phase transitions have been studied exclusively in spatial crystals: structures that are periodically modulated only in space but static in time (left inset of Fig. 1a)[7]. During a spatial topological phase transition, through spatial parametric tuning, the energy band gap ($\omega$-gap) closes and reopens, accompanied by changes in the spatial topological invariant (Fig. 1e)[8].

Recently, time has been introduced as a new degree of freedom in crystalline systems, giving rise to photonic time crystals (right inset of Fig. 1a), in which the refractive index varies periodically in time while remaining uniform in space[9-14]. In contrast to spatial photonic crystals, photonic time crystals exhibit momentum band gaps ($k$-gaps) in their band structures due to the breaking of time-translation symmetry, where light can be amplified rather than forbidden[13-18]. Nevertheless, similar to the spatial case, temporal parametric tuning triggers the closing and reopening of the $k$-gap, leading to a change in the temporal topological invariant and thus a temporal topological phase transition (Fig. 1f)[10,18,19]. Notably, in both conventional spatial and emerging temporal photonic crystals, periodic modulation is introduced in only one dimension: space or time. This raises a fundamental question: what happens when a structure is modulated periodically in both space and time? The answer lies in photonic spacetime crystals (PSTCs) (middle inset of Fig. 1a), where the space and time dimensions are intertwined, resulting not only in separated $\omega$- and $k$-gaps[20], but also in unique mixed energy-momentum band gaps ($\omega k$-gaps)[21-27]. Previous studies have primarily focused on the existence and excitation of specific spatiotemporal topological states[26,27], with experimental demonstrations limited to synthetic dimensions[20]. However, a unified theoretical framework for spatiotemporal topological phase transitions that describes the evolution of band-gap regimes and transformation between spatial and temporal topological characterizations remains elusive. Moreover, the exotic $\omega k$-gaps and spatiotemporal topologically localized states have not been observed in a physically realized PSTC constructed from realistic materials.

In this work, we theoretically propose and experimentally demonstrate spatiotemporal topological phase transitions in a genuine PSTC for the first time. We establish a unified framework for the emerging spatiotemporal topological physics, in which purely spatial and purely temporal topological phenomena emerge only as special cases. Using a dynamically modulated microwave transmission-line metamaterial, we systematically measure photonic band structures across varying spatiotemporal modulation depths, demonstrating the complete evolution of spatiotemporal topological phase transitions. We further realize a distinctive spatiotemporal topologically localized state and confirm its robustness against spatiotemporal disorder and its consistency with relativistic causality. These results validate the spatiotemporal bulk-boundary correspondence for PSTCs.

## Results

### Spatial, temporal, and spatiotemporal topological phase transitions

We consider a PSTC with spatial and temporal lattice periods $\Lambda$ and $T$, respectively. Its permittivity is independently modulated in space and time, taking the form $\varepsilon(x,t) = \varepsilon_{\mathrm{r}}[1+\delta_x\cos(G_0 x)][1+\delta_t\cos(\Omega_0 t)]$, where $\delta_x$ and $\delta_t$ denote the spatial and temporal modulation depths, respectively, and $G_0 = 2\pi/\Lambda$ and $\Omega_0 = 2\pi/T$ are the corresponding modulation frequencies. A typical distribution of $\varepsilon(x, t)$ within a unit cell is shown in Fig. 1b. Our analysis is based on the four-band coupled-mode theory, which has been used to predict the generation of topological event wave packets[27]. Here, we extend this framework to investigate the spatiotemporal topological phase transitions. Within this theoretical framework, the 0$^{\text{th}}$ and -1$^{\text{st}}$ harmonics of the spatial and temporal field distributions capture the forward-propagating, spatial Bragg-reflected, and temporal Bragg-reflected waves. These harmonics are sufficient to describe the $\omega k$-gap in the first Brillouin zone[27,28]. The field distribution can therefore be written as:

$$E(x,t) = \sum_{n,m} \psi_{(n,m)} \exp\{i[(k+nG_0)x - (\omega + m\Omega_0)t]\}, \tag{1}$$

where $(n, m) = (0, 0), (-1, 0), (0, -1)$ or $(-1, -1)$. We further derive the nonlinear eigenvalue equation for the PSTC (see Supplementary Section S1):

$$\boldsymbol{H}(k,\omega)\boldsymbol{\psi} = \begin{pmatrix} a_{(0,0)} & \frac{\delta_x}{2} & \frac{\delta_t}{2} & \frac{\delta_x\delta_t}{4} \\ \frac{\delta_x}{2} & a_{(-1,0)} & \frac{\delta_x\delta_t}{4} & \frac{\delta_t}{2} \\ \frac{\delta_t}{2} & \frac{\delta_x\delta_t}{4} & a_{(0,-1)} & \frac{\delta_x}{2} \\ \frac{\delta_x\delta_t}{4} & \frac{\delta_t}{2} & \frac{\delta_x}{2} & a_{(-1,-1)} \end{pmatrix} \begin{pmatrix} \psi_{(0,0)} \\ \psi_{(-1,0)} \\ \psi_{(0,-1)} \\ \psi_{(-1,-1)} \end{pmatrix} = 0, \tag{2}$$

where

$$a_{(n,m)}(k,\omega) = 1 - \frac{c^2(k+nG_0)^2}{\varepsilon_{\mathrm{r}}(\omega+m\Omega_0)^2}. \quad (3)$$

We find that the coupling matrix $\boldsymbol{H}(k,\omega)$ in Eq. (2) exhibits a high degree of symmetry, reflecting the interactions among field modes of different orders. Since $\delta_x$ and $\delta_t$ are associated with the strengths of spatial and temporal Bragg reflections[29], respectively, their corresponding terms appear exclusively in the off-diagonal elements. In particular, $\delta_x/2$ and $\delta_t/2$ represent the spatial and temporal coupling strengths, respectively, while their product, $\delta_x\delta_t/4$, denotes the spatiotemporal coupling strength. By solving $\det[\boldsymbol{H}(k,\omega)] = 0$, we obtain the photonic band structure relating $k$ and $\omega$, from which the $\omega$-, $k$-, and $\omega k$-gaps can be identified (see Supplementary Section S1 and Figs. 1e-h).

The spatial and temporal Zak phases provide a joint characterization of the topology of this PSTC system[20,26]. When an $\omega$-gap opens, the band forms a closed loop over the full momentum Brillouin zone, and its topology is characterized by the spatial Zak phase ($\gamma_s$). Conversely, when a $k$-gap opens, the band forms a closed loop over the full frequency Brillouin zone, and its topology is characterized by the temporal Zak phase ($\gamma_t$). In the mixed $\omega k$-gap, the two descriptions coexist as projections of the same spatiotemporal band structure. Spatial and temporal Zak phases are quantized to 0 or π, defining four distinct topological classes: $0_s$, $\pi_s$, $0_t$, and $\pi_t$, where the subscripts $s$ and $t$ specify the spatial and temporal sectors, respectively (see Supplementary Section S1). By introducing the dimensionless spatiotemporal aspect ratio $\eta = (\sqrt{\varepsilon_{\mathrm{r}}}\Lambda)/(cT) = (\sqrt{\varepsilon_{\mathrm{r}}}\Omega_0)/(cG_0)$[27], we construct the complete topological phase diagram of the PSTC, as shown in Figs. 1c,d. The regions in different colors correspond to different types of band gaps, separated by the boundaries $\eta^2(2 \pm \delta_x)(2 \pm \delta_t) = 4$, where the two ± signs are independent. The gray lines indicate the topological phase transition points.

Specifically, in the orange regions in Fig. 1c, the spatial coupling significantly exceeds the temporal coupling, allowing only an $\omega$-gap to open in the photonic band structure. Particularly, when the temporal modulation depth $\delta_t$ is zero, the spatiotemporal coupling strength $\delta_x\delta_t/4$ in Eq. (2) vanishes. Thus, the system reduces to a pure spatial photonic crystal, whose spatial topological phase transition is induced solely by varying the spatial modulation depth along the A1-A3 path in the phase diagram (Fig. 1c). The corresponding band structures are presented in Fig. 1e, with the calculation method detailed in Supplementary Section S2. Conversely, when the temporal coupling dominates (green regions in Fig. 1c), the system allows only a $k$-gap to open. With zero spatial modulation depth $\delta_x$ and zero spatiotemporal coupling strength $\delta_x\delta_t/4$, the system reduces to a pure photonic time crystal, whose temporal topological phase transition is controlled by varying the temporal modulation depth only (from points B1 to B3 in Fig. 1c), as shown in Fig. 1f. Note that purely spatial and purely temporal topological phase transitions require one of $\delta_x$ or $\delta_t$ to be zero, for which the spatiotemporal aspect ratio $\eta$ is undefined. However, when both $\delta_x$ and $\delta_t$ are small but nonzero, the $\omega$-gap and $k$-gap coexist and remain separated for $\eta \neq 1$, as shown in Fig. 1g and marked by the yellow regions in Figs. 1c,d. This type of band gap has previously been identified in synthetic dimensions[20], and its underlying mechanism has been theoretically elucidated[21,29]. In this regime, we further find that the $\omega$-gap and $k$-gap are independently controlled by $\delta_x$ and $\delta_t$, respectively (see Fig. 1g and the corresponding points C1 to C5 in Fig. 1d), and their band-gap types remain unchanged at a fixed $\eta$. Interestingly, this band structure exhibits a ring-like feature arising from spacetime folding, for which the Zak phase cannot be defined in either $k$ or $\omega$ parameter space.

When the spatial and temporal coupling strengths are relatively large and comparable (blue regions in Fig. 1c), both the energy and momentum band gaps open simultaneously at the same position, resulting in a mixed $\omega k$-gap. Taking $\eta = 1$ as a representative example, as the spatial ($\delta_x$) and temporal ($\delta_t$) modulation depths vary from points D1 to D9 in Fig. 1c, we find the band structures and their spatial or temporal Zak phases evolve complexly, revealing a complete mixed spatiotemporal topological phase transition, as shown in Fig. 1h. Specifically, from D1 to D3, two energy bands (orange curves in panel D1 of Fig. 1h) converge, degenerate, and then open a pure $k$-gap. During this transition, their Zak phases change from $\pi_s$ in the spatial type to $0_t$ in the temporal type. From D3 to D5, this $k$-gap closes and reopens a different $\omega k$-gap, and the Zak phases change from $0_t$ to $0_s$. From D5 to D7, the two momentum bands (green curves in panel D5 of Fig. 1h) become degenerate and then open a pure $\omega$-gap, with their Zak phases changing from $\pi_t$ to $0_s$. Finally, from D7 to D9, this $\omega$-gap closes and reopens again as a new $\omega k$-gap, and the Zak phases change from $0_s$ to $0_t$. As a result, the mixed spatiotemporal topological phase transition involves changes not only in the values of the topological invariants but also, notably, in their types and in the band-gap type.

## Observation of band-gap type evolution

In our experiments, we focus on the mixed spatiotemporal topological phase transitions, which are of particular interest because they involve changes in band-gap types. We employ a dynamically modulated microwave spoof surface plasmon polariton (SSPP)[30,31] transmission line fabricated on a printed circuit board (PCB) to construct the PSTC (Fig. 2a). The SSPP platform offers two advantages. First, its strong dispersion ensures that only four bands appear under spatiotemporal modulation (see Supplementary Section S3 and Fig. S1), consistent with the four-band coupled-mode theory described above. Second, as demonstrated in both theory[28] and experiment[32], the surface plasmon resonance can significantly expand the momentum band gap, thereby relaxing the requirement for a high temporal modulation depth.

Figures 2a,b present a photograph of the experimental sample and a schematic of the unit cell. The top copper layer is patterned with alternating long and short slots, thereby introducing spatial periodic modulation and facilitating the emergence of spatial topology[33]. Each copper strip is connected to a varactor diode on the bottom layer through a via. A time-varying, periodic signal is applied to the varactors through a DC-block bandpass filter, resulting in a periodic temporal modulation. A DC bias voltage of $V_{\mathrm{DC}}$ = 15 V is applied to the varactors, enabling their average capacitance of approximately 10.16 pF (different from the transmission line's equivalent static distributed capacitance $C_0$). The spatial lattice constant $\Lambda$ is 8 mm, with air slot lengths $d_1$ = 17.5 mm and $d_2$ = 22.5 mm, $w_1 = w_2$ = 2 mm, and the gap $g$ = 1 mm, respectively (see Supplementary Section S4, Figs. S2 and S3 for more details).

Figure 2c shows the equivalent circuit model of the dynamically modulated SSPP transmission line. The short and long slots in the top copper layer are modeled as inductors $L_1$ and $L_2$, respectively, and the mutual inductances ($M_1$, $M_2$, $M_3$, ...) account for magnetic field coupling between adjacent slots (see Method and Supplementary Section S5). The fitted values are $L_1$ = 20.5 nH, $L_2$ = 24.7 nH, and $M_n = 8.0/n^{3.2}$ nH, where $n$ is a positive integer (see Supplementary Section S6). The time-varying effective distributed capacitance $C_{\mathrm{eff}}(t)$ incorporates a gap capacitance $C_{\mathrm{g}}$, the varactor capacitance $C_{\mathrm{V}}(t)$, the filter's influence, and the amplifier's output resistance, whose equivalent circuit is shown in the lower-left inset of Fig. 2c. Under temporal modulation, $C_{\mathrm{eff}}(t)$ oscillates about the central value $C_0$ = 19.3 pF with a cosine-like function of period $T$ and amplitude $\Delta C$, as depicted in the lower-right inset of Fig. 2c. In the experimental measurements, we use an arbitrary waveform generator (AWG) to generate a cosine modulation signal with a period of $T$ = 1.517 ns (659 MHz), which satisfies $\eta$ = 1. This modulation signal is then split, amplified, and applied to the varactors to generate the time-modulated $C_{\text{eff}}(t)$. A Gaussian signal with a center frequency of 329.5 MHz from another channel of the AWG is injected into one end of the SSPP transmission line via a custom excitation source. The magnetic field distributions of the propagating wave are measured using a magnetic probe, then amplified and recorded by an oscilloscope (see Method and Supplementary Section S7).

We define the spatial and temporal modulation depths as $\Delta L/L_0$ and $\Delta C/C_0$, respectively, where $L_0 = (L_1 + L_2)/2$ is the average inductance and $\Delta L = (L_2 - L_1)/2$ is the inductance modulation amplitude. Based on the equivalent circuit model in Fig. 2c, we calculate the phase diagrams of the SSPP-based PSTC at different spatial and temporal modulation depths with $\eta = 1$, as shown in Fig. 2d. Figure 2e further illustrates a detailed view of the red dashed box in Fig. 2d, with the band-gap width indicated by the color scale. Specifically, the spatial ($\omega$-gap, orange region) and temporal ($k$-gap, green region) topological phases are distinguished by their respective normalized band-gap widths $\Delta\omega$ and $\Delta k$, while the spatiotemporal topological phase ($\omega k$-gap, blue region) is represented by its $\omega k$-gap width $\min(\Delta\omega, \Delta k)$.

At a fixed spatial modulation depth ($\Delta L/L_0 = 0.093$), we measured the photonic band structures of the SSPP-based PSTC under five different temporal modulation depths (points I to V in Fig. 2e) by varying the modulation voltage amplitude, as presented in Fig. 2f-j (see Supplementary Sections S8 and S9). At a low temporal modulation depth (Fig. 2f, $\Delta C/C_0 = 0.1$), the measured (color map) and calculated (blue dotted lines) band structures exhibit an $\omega$-gap at $\omega = \pi/T$, within which the wave transmission is prohibited. The gray dotted lines depict the full band structures folded from the blue dotted lines due to the periodic spatiotemporal modulation. As the temporal modulation depth increases, two of the energy bands undergo degeneracy (Fig. 2g, $\Delta C/C_0 = 0.148$) and then reopen to form an $\omega k$-gap at moderate temporal modulation depth at $\omega = \pi/T$ and $k_x = \pi/\Lambda$ (Fig. 2h, $\Delta C/C_0 = 0.18$). In this $\omega k$-gap, wave amplification — a key signature of the band gap in momentum[13,18] — occurs and manifests as a bright spot. Because the photonic band structure was measured using unidirectionally propagating waves, the observed $\omega k$-gap is asymmetric, with amplification occurring only for $k_x < \pi/\Lambda$. As the temporal modulation depth further increases, the other two energy bands degenerate (Fig. 2i, $\Delta C/C_0 = 0.217$), and finally reopen to form a pure $k$-gap at $k_x = \pi/\Lambda$ under high temporal modulation depth (Fig. 2j, $\Delta C/C_0 = 0.25$), where wave amplification causes a brighter spot in the whole $k$-gap. Note that the bright stripes at $\omega = \pi/T$ in Figs. 2f-j are attributed to the time-modulation drive signal. This band-gap type evolution indicates transitions in the types of topological invariants, corresponding to spatiotemporal topological phase transitions between purely spatial (Fig. 2f), purely temporal (Fig. 2j), and mixed spatiotemporal (Fig. 2h) topological phases.

**Observation of spatial, temporal, and spatiotemporal topological states**

In PSTCs, spatial, temporal, and spatiotemporal topological states can emerge at interfaces between structures with different spatiotemporal modulations. In the experiment, we fabricated an SSPP-based PSTC sample with a spatial interface, where the left and right halves consist of different spatial unit cells with two mirror-symmetric air-slot configurations, respectively, as shown in Fig. 3a. By sequentially applying two temporal modulations of equal magnitude but opposite signs (i.e., opposite phases) to the sample, four distinct types of spatiotemporal unit cells can be created, denoted as PSTC1–PSTC4, as shown in Fig. 3b. In the spatiotemporal domain, these four configurations form a spatial interface at $x = 0$ and a temporal interface at $t = 0$, indicated by the orange and green dashed lines, respectively, in Figs. 3c-e. On either side of the spatial (or temporal) interface, the spatial (or temporal) modulation has equal magnitude but opposite signs. A Gaussian source (yellow star) is placed near the spatial interface to excite the potential topological states (see Supplementary Section S8).

We measured the instantaneous evolution of the magnetic field distribution, as shown in Figs. 3c-e. For the $\omega$-gap case (Fig. 3c), the PSTCs have different band-gap spatial topological invariants $\nu_s$ across the spatial interface, whereas the temporal topological invariant $\nu_t$ of the band gap is undefined on both sides of the temporal interface because the band structure contains only a pure $\omega$-gap (see Supplementary Section S5 for the definition of band-gap topological invariants). The spatial interface is therefore nontrivial, while the temporal interface is trivial, resulting in a field localized at the spatial interface and decaying in time. The corresponding time- and space-averaged magnetic-field profiles, $\langle|H_z|\rangle_t(x)$ and $\langle|H_z|\rangle_x(t)$ (top and right insets in Fig. 3c), confirm the exponential spatial localization and temporal decay, with the blue dashed lines showing exponential fits to the field maxima. For the $k$-gap case (Fig. 3d), the spatial and temporal interfaces become trivial and nontrivial, respectively. Accordingly, the field energy is spatially delocalized, as shown in the top inset of Fig. 3d, where evident spatial bulk states emerge. In contrast, the right inset shows that the field undergoes exponential amplification before the temporal interface and exponential decay afterward, indicating temporal localization at the interface and the formation of a temporal topological state. Notably, the averaged magnetic field in the right inset of Fig. 3d increases again long after the time interface due to parametric instability within the *k*-gap, consistent with a previous report in a purely time crystal[18]. For the $\omega k$-gap case (Fig. 3e), both the spatial and temporal interfaces are nontrivial, confining the field to their intersection and forming a spatiotemporal topological state (or named spatiotemporal topological event[20]). The averaged magnetic-field profiles (top and right insets) demonstrate simultaneous exponential localization in space and time. These spatial, temporal, and spatiotemporal topological states are consistently predicted by the topological invariants, confirming the validity of bulk-boundary correspondence in PSTCs.

**Robustness and causality of the spatiotemporal topological events**

Finally, we experimentally verify the topologically protected robustness and relativistic causality of the spatiotemporal topological states in PSTCs. These properties have been observed in a synthetic dimension[20] but have yet to be confirmed in physically realized PSTCs. We introduce disordered spatiotemporal modulation depths near the spatiotemporal interfaces or within the bulk of the SSPP-based PSTCs, respectively, as illustrated in the left and lower insets of Figs. 4a,b (see Supplementary Fig. S3 for photographs of the fabricated samples). In both cases, the measured magnetic field intensity distributions of the spatiotemporal topological states exhibit robustness to spatiotemporal disorder, confirming their nontrivial topological properties. More interestingly, the occurrence of spatiotemporal topologically localized events is constrained by relativistic causality. That is, a localized event can be triggered only if the source lies within its past light cone (represented by the grey dashed lines). As shown in Fig. 4c, the past light cone is centered at the intersection of the spatial and temporal interfaces, with its speed of light determined by the low-frequency limit of the SSPP dispersion (see Supplementary Section S10). A spatiotemporal topological event occurs when a point source (yellow star) lies within this light cone. In contrast, when the source is placed outside this past light cone, the excitation cannot causally affect the intersection point, even though its absolute time precedes the temporal interface (green dashed line). Consequently, such a localized event cannot occur, as illustrated in Fig. 4d.

**Conclusion**

In summary, we have theoretically predicted and experimentally demonstrated a spatiotemporal topological phase transition in a genuine PSTC, which unifies purely spatial, purely temporal, spatially and temporally separated, and mixed spatiotemporal topological phases in a single system. Using a physically realized PSTC based on a microwave SSPP transmission line, we realized exotic transitions between different band-gap types, including $\omega$-, $k$-, and mixed $\omega k$-gaps, which are absent in conventional photonic crystal systems. Furthermore, we directly observed spatial, temporal, and spatiotemporal topological states at interfaces among four PSTCs with distinct topological invariants, consistent with the spatiotemporal bulk-boundary correspondence. These spatiotemporal topological states exhibit robustness against spatiotemporal disorder and are strictly governed by relativistic causality. These novel spatiotemporal topological phases and phase transitions are anticipated to enable promising applications in tunable topological lasers[34,35], highly sensitive optical detection, next-generation wireless communication[36], and dynamic radar systems. The presented concepts are envisioned to be extended to other physical disciplines, such as condensed matter physics, acoustics, phononics, thermotics, and electrical circuits.

**Data availability**

The data that support the findings of this study are available from the corresponding author upon reasonable request.

**Acknowledgments**

Z.G. acknowledges funding from the National Key R&D Program of China (grant No. 2025YFA1412300), National Natural Science Foundation of China (grant no. 62361166627 and 62375118), Guangdong Basic and Applied Basic Research Foundation (grant No. 2024A1515012770), Shenzhen Science and Technology Innovation Commission (grant No. 20230802205352003), and High-level Special Funds (grant No. G03034K004). Z.B.Zhu acknowledges the funding from the China Postdoctoral Science Foundation (grant No. 2025M773439). Y.M. acknowledges the funding from the National Natural Science Foundation of China (grant No. 12304484) and the Guangdong Basic and Applied Basic Research Foundation (grant No. 2024A1515011371). Z.X.Zhu acknowledges the funding from the National Natural

Science Foundation of China (grant No. 12404053) and the Guangdong Basic and Applied Basic Research Foundation (grant No. 2025A1515012284). X.X. acknowledges the funding from the National Natural Science Foundation of China (grant No. 62405053), the Basic and Applied Basic Research Foundation of Guangdong Province (grant No. 2025A1515012229), and the Youth S&T Talent Support Program of Guangdong Provincial Association for Science and Technology (grant No. SKXRC2025440).

**Author Contributions**

Z.G. initiated and supervised the project; Z.B.Zhu developed the theory, performed numerical calculations, designed and built the experimental setup; Z.B.Zhu, B.H., and S.X. performed the experimental measurements, with contributions from Z.X.Zhu; Z.B.Zhu and B.H. analyzed data; J.C., Y.M., and X.X. participated in discussions of this project; Z.B.Zhu and Z.G. wrote and revised the manuscript.

**Competing Interests**

The authors declare no competing interests.

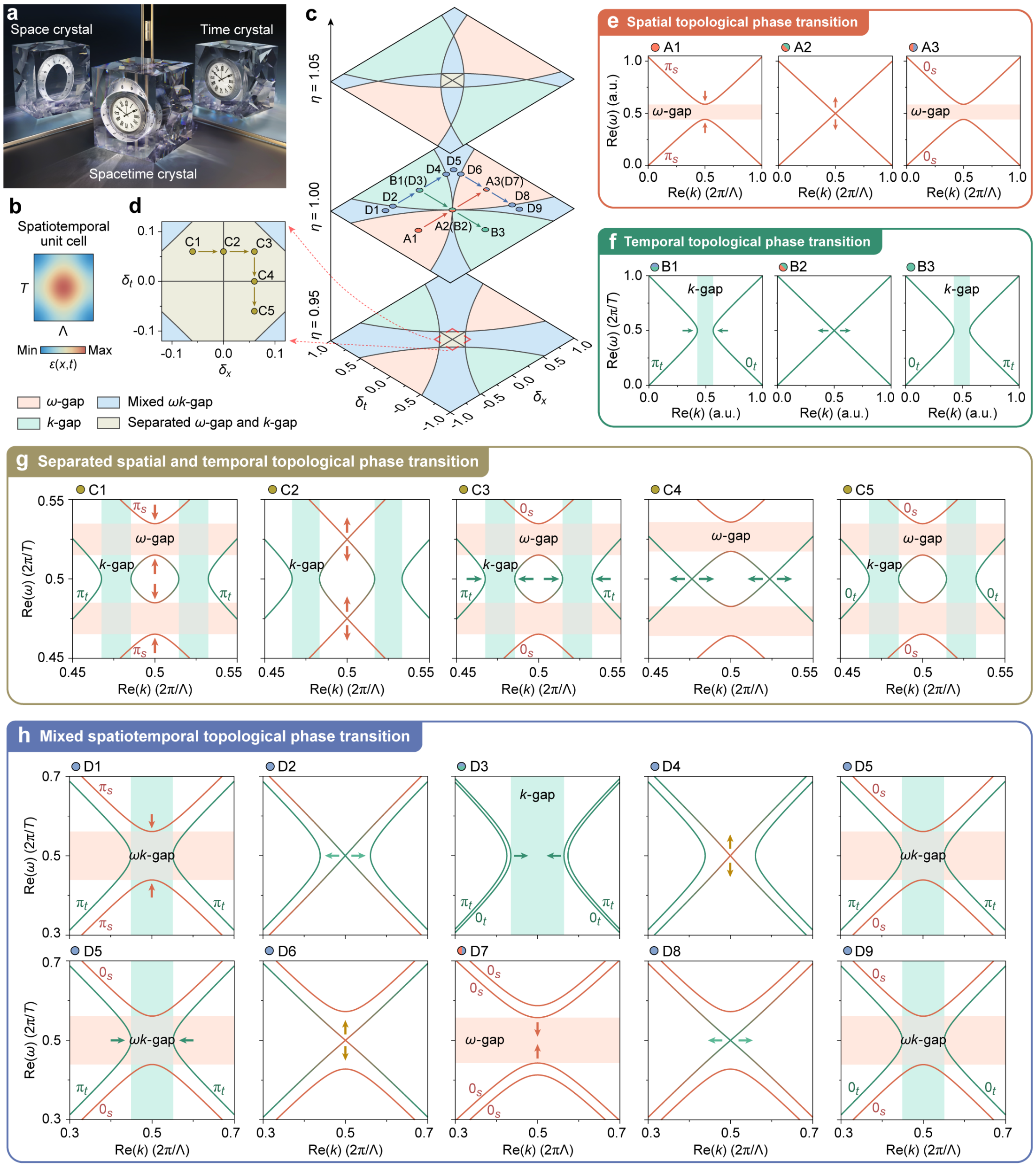

**Fig. 1 | Spatial, temporal, and spatiotemporal topological phase transitions. a** Schematic of the space crystals, time crystals, and spacetime crystals, the first two are only special cases of the last one. **b** A typical spatiotemporal distribution of dielectric constant in a unit cell, defined as $\varepsilon(x,t) = \varepsilon_r[1 + \delta_x\cos(2\pi x/\Lambda)][1 + \delta_t\cos(2\pi t/T)]$, where $\delta_x$ and $\delta_t$ represent the spatial and temporal modulation depths, respectively; $\Lambda$ and $T$ are the spatial and temporal lattice constants, respectively. **c**,**d** Phase diagram of a photonic spacetime crystal as a function of $\delta_x$, $\delta_t$, and $\eta$, where $\eta = (\sqrt{\varepsilon_r}\Lambda)/(cT)$ is the spatiotemporal aspect ratio. Different colored regions represent different types of band gaps, and grey lines indicate the topological phase transition points. **e-h** Evolution of purely spatial (**e**), purely temporal (**f**), separated spatial and temporal (**g**), and mixed spatiotemporal (**h**) topological phase transitions with varying $\delta_x$ and $\delta_t$ corresponding to the colored dots in (**c**) or (**d**). Arrows in (**e**)-(**h**) indicate the moving trends of photonic bands. $0_s$, $\pi_s$, $0_t$ and $\pi_t$ denote the topological invariants (Zak phase) of the photonic bands.

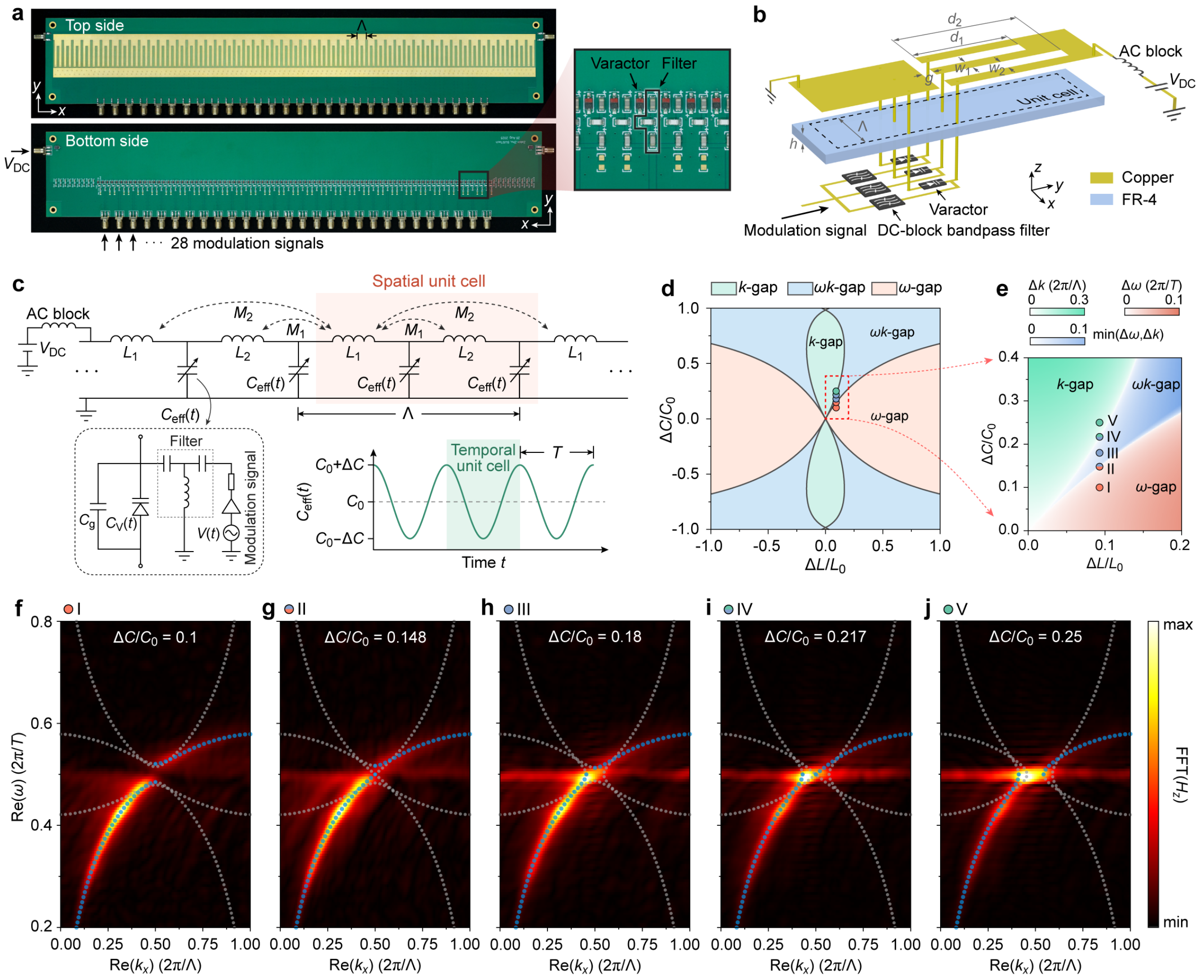

**Fig. 2 | Experimental demonstration of band-gap type evolution in mixed spatiotemporal topological phase transitions. a** Photograph of the fabricated PSTC based on a dynamically modulated SSPP transmission line. The right inset shows the enlarged photograph of the varactors and filters. **b** Schematic diagram of a spatial unit cell. **c** Equivalent circuit model of the time-modulated SSPP transmission line. **d** Analytically calculated phase diagram of the PSTC, where $\Delta L/L_0$ and $\Delta C/C_0$ denote the spatial and temporal modulation depths, respectively. Different colored regions represent different types of band gaps. **e** Enlarged view of the region enclosed by the red dashed box in (**d**). Color scales in (**e**) represent the band-gap width. **f-j** Measured (color map) and calculated (blue dotted line) photonic band structures with unidirectional wave propagation corresponding to the colored dots I ~ V in (**e**). The gray dotted lines represent the full band structures folded from the blue dotted lines due to periodic spatiotemporal modulation.

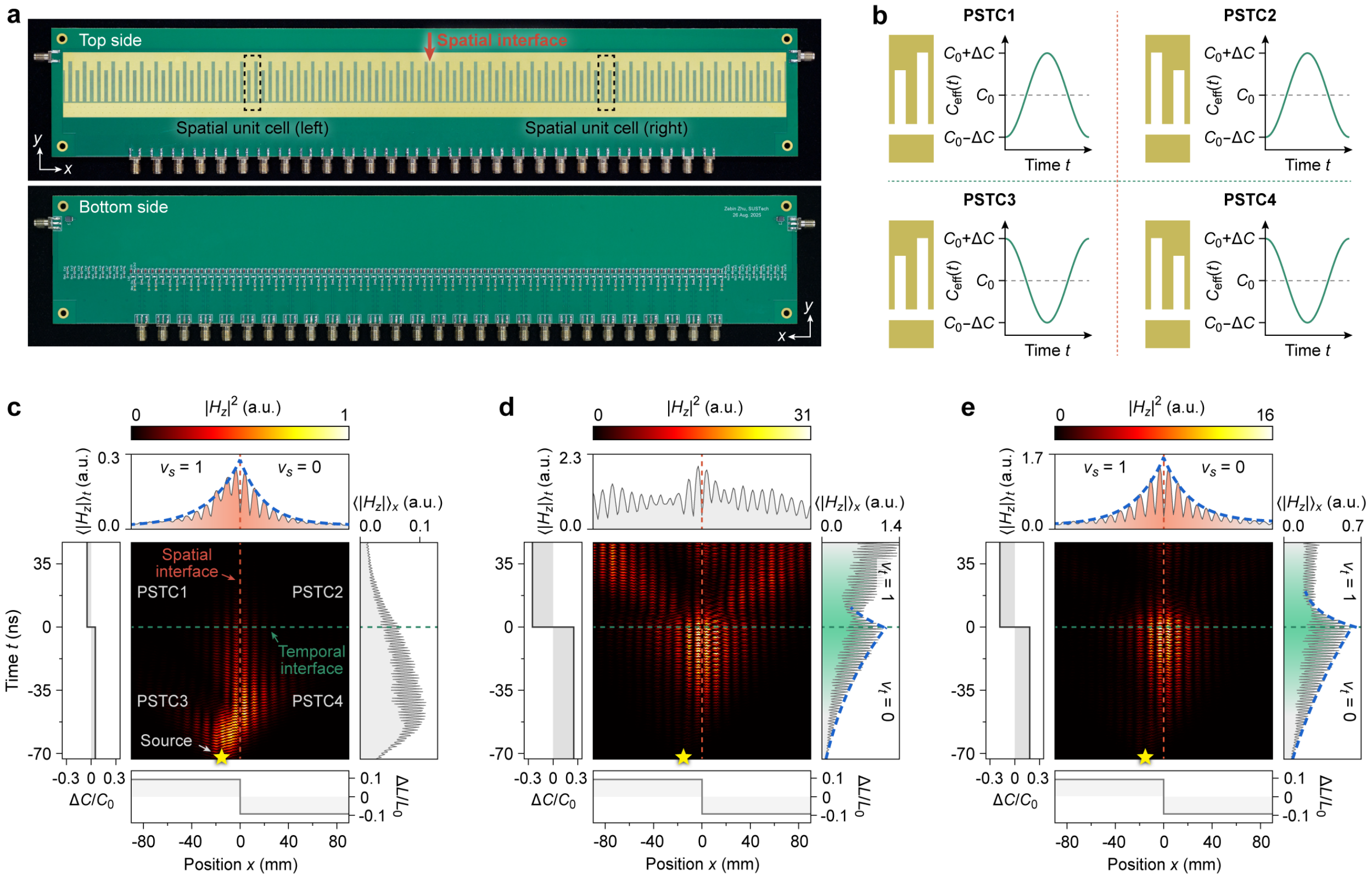

**Fig. 3 | Observation of spatial, temporal, and spatiotemporal topological states. a** Photograph of the sample used for topological-state excitation. It contains a spatial interface with mirror-symmetric spatial unit cells on both sides. **b** Schematics of different spatiotemporal unit cells. By applying either of two temporal modulations with equal magnitude but opposite phases to the sample, four distinct types of spatiotemporal unit cells can be created, denoted as PSTC1–PSTC4. **c-e** Experimentally measured magnetic-field distributions of spatial [(**c**), $|\Delta C/C_0|$ = 0.05], temporal [(**d**), $|\Delta C/C_0|$ = 0.25], and spatiotemporal [(**e**), $|\Delta C/C_0|$ = 0.18] topological localized states, with $|\Delta L/L_0|$ fixed at 0.093. In (**c**)-(**e**), the orange and green dashed lines indicate the spatial and temporal interfaces, respectively; the yellow stars indicate the point sources; the left and bottom insets show the temporal and spatial modulation depths as functions of time and space, respectively; the top and right insets depict the temporal and spatial averages of the magnetic-field magnitude, respectively; the blue dashed curves are exponential fits to the peak values of the averaged magnetic-field magnitude; $v_s$ and $v_t$ are spatial and temporal topological invariants of band gaps, respectively.

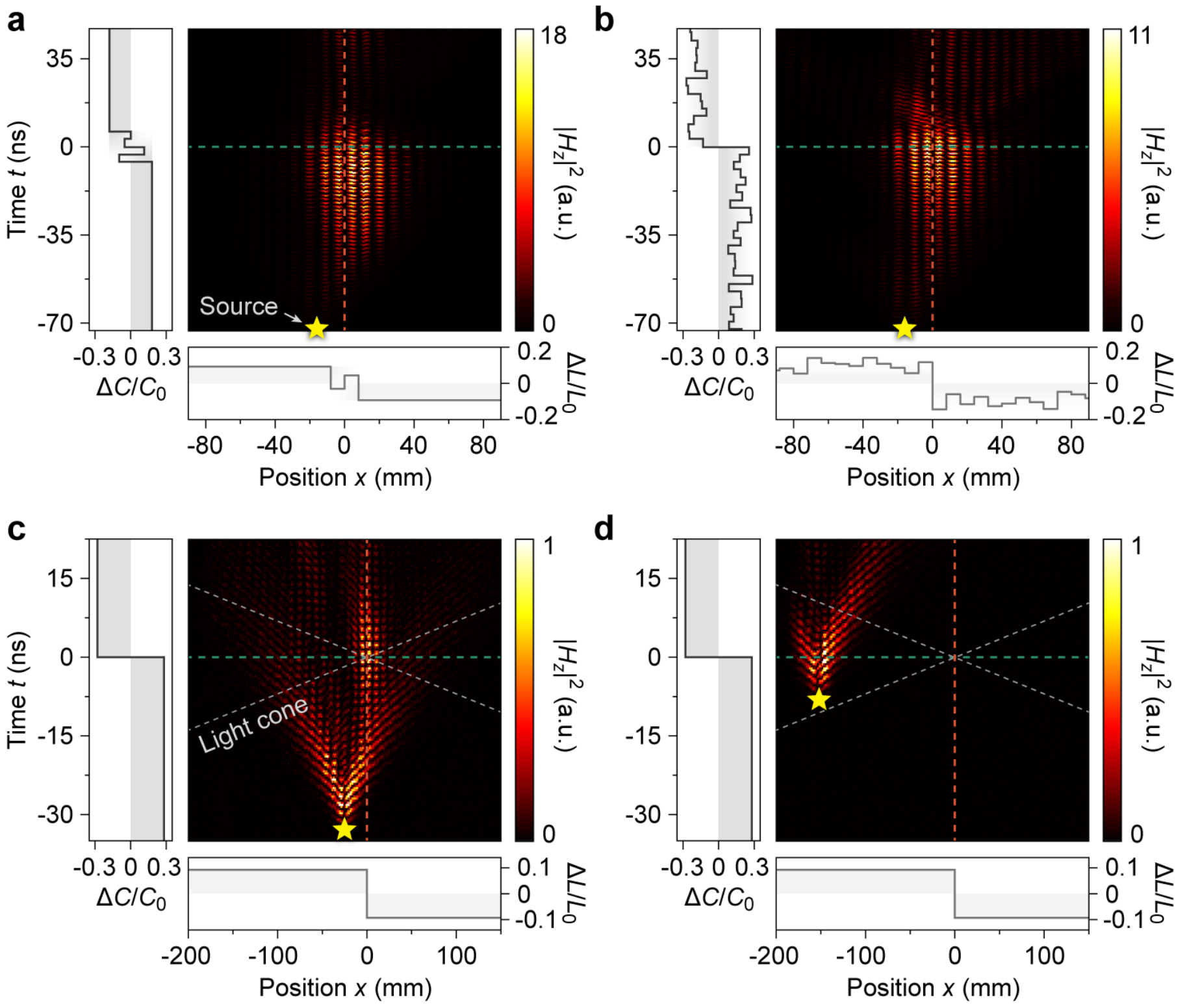

**Fig. 4 | Robustness and causality-governed excitation of spatiotemporal topological states in a genuine PSTC. a**,**b** The spatiotemporal topological state remains robust against spatiotemporal disorders near the spatial (orange dashed line) and temporal (green dashed line) interfaces (**a**) and in the bulk (**b**), respectively. The left and lower insets represent the disorders of temporal and spatial modulation depths. **c**,**d** The excitation of a spatiotemporal topological state is possible only if the source (yellow star) excitation lies within its past light cone (**c**), and is causally impossible if the excitation lies outside it (**d**). The grey dashed lines in (**c**) and (**d**) indicate the light cones of the SSPP waveguide.